# Stratégies d'automédiation : de l'expression de soi au jeu des intersubjectivités

## Etude de la représentation de l'usager dans Livejournal et Touchgraph

**Fanny Georges**

*Centre de recherche images et cognition (CRIC) Paris I*
*47 - 53, rue des bergers*
*75015 PARIS.*
*fannygeorges@free.fr*

RÉSUMÉ. *Acteur de sa présentation et de sa représentation en ligne, le diariste dessine les contours de son existence diégétique en élaborant une stratégie d'automédiation. La représentation de soi est une création personnelle déterminée par l'interface et les fonctionnalités du logiciel. Son usage manifeste des stratégies disparates, de l'épanchement solipsiste à la collection de tests et plaisanteries destinés à briguer le « top 10 » des blogs les plus visités. Le logiciel incite l'usager à sortir de l'espace solitaire de l'écriture pour s'engager dans l'espace communautaire de la publication. Une analyse pragmatique des enjeux de la représentation de soi dans le blog Livejournal et son navigateur Touchgraph est l'occasion d'observer comment l'intimité se prête au jeu des intersubjectivités.*

ABSTRACT. *Actor of its presentation and actor of its online representation, the diarist draws his diegetic existence by setting up a strategy of automediation. The Self-representation is a personal creation determined by the interface and the functionalities of the software. A pragmatic approach of the Self-representation in the Livejournal Blog and the Touchgraph Livejournal browser provides a way to observe the play between intimacy and intersubjectivity.The software leads the user from the lonely space of writing to the community space of publication.*

MOTS-CLÉS : *avatar, représentation de soi, expression, visualisation, identité, réflexivité.*

**1. Introduction**

En coprésence, le sujet communique par sa posture, ses vêtements, avant même de verbaliser son message [GOF 73]. Le corps émet des signes naturels [ECO 88]. Volontaire ou non, la communication est inévitable. A distance, l'interface graphique s'interpose. Les signes comportementaux y sont transposés par l'usager qui s'adapte à elle. Dans la famille des logiciels de communication dont fait partie le cyberjournal, l'usager n'existe pas sans sa représentation. Son inscription[1] est prise d'existence.

Indicielle, la représentation de l'usager est la trace d'une action [VER 97] sur le support. Signe actuel de présence au monde, elle est aussi l'empreinte de l'individualité de son acteur. De même que le registre de langue qualifie à la fois le locuteur et reflète la manière dont il perçoit le destinataire, de même l'ensemble des signes audio-scripto-visuels manifestent la posture et l'intentionnalité de l'émetteur[2]. Nous désignerons par *représentation de l'usager*[3] l'ensemble de ces phénomènes perçus par un tiers comme constituant la représentation d'un individu, qu'ils relèvent de l'automédiation ou de la communication interpersonnelle. Les penser sous un même concept pose un cadre à l'observation de la stratégie de médiation de soi dans un contexte communautaire. [BIC 98]

Avec le développement du web dynamique, les journaux intimes sur internet (*Ublog*[4], *Livejournal*[5], *20six*[6]) ont pris la relève des pages personnelles ou homepages [KLE 02] dans les pratiques d'automédiation[7]. En janvier 2005, PEW[8]

---

1. « Inscription » peut être entendu ici à la fois au sens d'inscription graphique de la présence et au sens d'inscription au service en remplissant une fiche d'inscription, le premier étant conditionné par le second.
2. Dominique Combe [COM 94] se réfère à Merleau Ponty pour proposer une stylistique phénoménologique qui « devrait s'appuyer, donc, sur l'indistinction de la pensée et du langage dans l'acte de style, qui élabore une vision du monde en même temps que l'écriture se cherche. » (p.89) Dans le domaine des nouvelles technologies de la communication, ce monisme serait inhérent à l'univers, tout enclos en l'interface, de la diégèse communicationnelle, tandis que du point de vue de l'émetteur comme du récepteur, le procès d'interprétation et la conscience du jeu des intersubjectivités contraindrait à conserver une approche dualiste de la pensée et de l'expression.
3. Nous réserverons l'emploi de représentation de soi à la désignation des signes reconnus par l'usager lui-même comme appartenant à sa représentation.
4. http://www.figureconcord.com/ublog/
5. http://www.livejournal.com/
6 http://www.20six.co.uk/
7 « les productions culturelles qui naissent de l'expression de soi, de la présentation de soi à travers ses goûts, ses passions sur le réseau » déploient une « culture sur mesure ou self culture ». Elles manifesteraient une « réversibilité systématisée des rôles entre auteur et rapporte que 27% des adultes américains utilisateurs d'internet ont consulté un blog. Initialement développés et mis en ligne par les utilisateurs même ou des membres de leur entourage, ils ont pris le pas sur les pages html en raison de la facilité de publication. Des sites ont proposé des solutions clé en main avec création d'un « site en trois clics » [ALL 03] hébergement, forums, newsletters, moteurs de recherche. C'est ainsi que la physionomie à présent si connue du blog s'est imposée, au point de tendre à ne désigner que les pages dynamiques comportant un calendrier, des messages datés (posts) et une liste de blogs favoris, là où la page personnelle aux cadres élastiques se caractérisait par une mise en page rudimentaire et des images mal optimisées pour le web.

Le rassemblement en communauté caractéristique du cyber-journal, est la marque d'une réelle volonté de l'exposition de soi et marque le passage du journal intime au journal intime-public [CAU 03]. Outil d'expression personnelle, le blog intègre des fonctionnalités de référencement interne : la première promesse à l'utilisateur est la publication, la seconde la visibilité intrinsèquement liée à la relation intersubjective [BIC 98]. Live journal est un blog américain basé sur un logiciel open source racheté début 2005 par *Sixapart*. La communauté, d'ampleur internationale, rassemble plus de 6 millions de comptes dont 70% sont ouverts par des femmes. Les membres publient des messages dans un journal personnel, collectif et des commentaires dans les journaux d'autres usagers[9].

L'arrivée des blogs audio (*Audioblog*[10]), l'adaptation du format au blog mobile[11] et la familiarisation croissante des utilisateurs avec la programmation font évoluer les pratiques. Cette mutation rapide des services fait du cyberjournal et des outils de communication en général des objets d'étude provisoires. Ils changent avec les pratiques technologiques. Parallèlement aux journaux personnels se sont développés les sites de réseaux sociaux (*networking social*) qui proposent notamment aux usagers de visualiser leur réseau de connaissances professionnelles (*Linkedin*[12]) ou amicales (*Friendster*[13]) dans le dessein de saisir des opportunités fructueuses de rencontre. Les blogs et logiciels de networking social relèvent encore à l'heure actuelle d'usages et de catégories distinctes. L'application *Touchgraph*[14], développée par Google, ne fait pas partie de l'offre de service initiale de *Livejournal*

---

lecteur-spectateur », produisant « un court-circuit de la différenciation, héritée de la modernité esthétique, des rôles culturels au profit d'une logique d'automédiation. » [ALL 02 :3]
8. http://pewinternet.org/pdfs/PIP_blogging_data.pdf
9. Un même compte peut recevoir plusieurs journaux individuels et collectifs. La publication peut se faire par le navigateur en ligne ou des logiciels clients en local téléchargeables comme Semagic.
10. http://www.audioblog.com/
11. Ou moblog, actualisable depuis un mobile comme Typepad (http://www.typepad.com/).
12. https://www.linkedin.com/home?trk=logo
13. http://www.friendster.com/
14. http://www.touchgraph.com/

et ne semble pas revendiquée par ses diaristes, mais permet de visualiser certaines informations utilisateur sous forme de graphe, de naviguer dans les réseaux et d'expérimenter différentes manières d'ordonner les éléments à l'écran. Ces deux logiciels complémentaires fourniront le terrain d'une analyse sémiopragmatique de la représentation de soi.

Après avoir relevé et classé les signes de l'interface qui représentent le diariste, nous réfléchirons à la façon dont s'organisent les différentes couches narratives qui se superposent en la pratique du service. Acteur de son identité diégétique, le diariste a à sa disposition un certain nombre de fonctionnalités qu'il peut utiliser pour gérer sa visibilité et dessiner ainsi les contours d'une intimité face à soi-même et face aux autres.

De la phase de création à la relation intersubjective tissée autour de l'objet créé intervient la problématique de la visibilité et de la diffusion. Or, tous les lecteurs étant également auteurs, la stratégie de diffusion deviendrait-elle une forme de jeu communautaire, celui d'être lu pour accroître son lectorat et ses statistiques ? Nous tenterons de discerner les stratégies d'action possible et les mettrons en regard avec la promesse initiale du service, s'exprimer soi-même.

## 2. La représentation de soi sous le regard du système

Le recours à la notion d'automédiation ou de médiation de soi nécessite de poser la question préalable de l'identité. Quelle est cette entité que l'émetteur transmet dans un mouvement apparemment réflexif ? Si l'usager et son avatar sont identiques dans la mesure où ils sont deux manifestations d'une même personne, l'avatar est bel et bien le personnage d'une diégèse[15] dont la construction s'élabore au fur à mesure de l'interaction.

### 2.1 Identités

Selon les théories de l'identité développées par Hume, le moi n'existe pas. L'identité recouvrirait une vérité nébuleuse, construction de l'imaginaire entre des souvenirs disparates – une représentation de soi tissée entre des souvenirs qui font sens par leur inclusion dans une diégèse identitaire. Or, l'expérience du personnage fait partie de l'existence de la personne, et l'identité du personnage comme représentation se tisse elle-même au fil de son élaboration.

La représentation de soi contient des informations dont certaines se rapportent au réel, d'autres à l'univers diégétique. Nous avons tenté, dans la figure 1, de les

---

15. Jacques Fontanille, à propos de la diégèse dans le contexte de l'hypertexte fictionnel : « La diégèse devient un nuage de possibles narratifs, au lieu d'une chaîne logiquement réglée à partir de la fin. » [FON 99] classer selon leur localisation dans l'espace communautaire et selon qu'elles relèveraient de l'identité personnelle, de l'identité diégétique, ou encore de la catégorie des identifiants.

L'identité personnelle est attachée au vécu quotidien. La fiction identitaire qui la produit fait intervenir les éléments vécus depuis sa naissance parmi lesquels les expériences identitaires virtuelles. Par simplification peut-être excessive, nous la différencions ici de l'identité diégétique, tout en sachant, comme nous l'avons déjà souligné, que cette dernière est incluse dans le procès d'unification des informations identitaires.

L'identité diégétique est l'identité du personnage qui s'exprime dans le blog à la première personne. Elle prend la forme, par exemple, du pseudonyme et de l'icone, qui représentent l'utilisateur dans les espaces de syndication[16] des blogs qui l'ont placé en favori.

Les identifiants relèvent de l'identité numérique. Points de passage entre le monde réel et le monde diégétique, ils permettent l'identification de l'acteur sur le réseau à laquelle sont associés l'IP, mais aussi le nom, l'adresse, les coordonnées bancaires si elles sont délivrées. Elles permettent le traçage. Information unique au sein d'une même communauté, l'identifiant ne peut être utilisé par deux membres[17], le mot de passe est une donnée confidentielle.

La représentation de soi dans Livejournal se constitue d'une partie publique et du back-office, contenant des informations privées[18]. La partie publique contient deux volets principaux, la page de profil et le journal. Le premier effectue la médiation entre le soi du monde réel et le soi du journal[19] ; le second contient les informations qui participent à la construction de la représentation de l'univers diégétique.

La figure 1 montre que des informations relevant explicitement de l'identité personnelle apparaissent sur la page de profil (âge, sexe, ville). Plus indirectement, la date de dernière mise à jour ou encore la datation des posts dans la partie journal ne font pas nécessairement sens dans la diégèse. Elles sont des signes extra-diégétiques, indices des moments de l'actualisation.

Les posts, que nous appréhendons ici comme signe unique relevant de l'univers diégétique en tant qu'ils en sont la semence même, renferment par ailleurs de nombreuses informations qui caractérisent la personne : l'orthographe, la syntaxe, le niveau de langue, la qualité de la narration, sont indices du contexte social dans lequel s'est formée l'identité personnelle. La qualité de la mise en page informe sur les compétences de l'usager en matière de développement, de sorte à ce qu'il soit par exemple aisé d'identifier un webmaster à la dextérité qu'il manifeste.

---

16. On parle en général de "syndication de contenu" pour désigner la possibilité de publier sur un site Web de façon automatisée du contenu provenant d'un autre site Web.
17. A la différence du *pseudonyme*, que l'utilisateur est libre de changer sans être contraint de changer de compte.
18. Nous ne nous attacherons qu'à détailler les éléments de la partie publique dans la mesure où ce sont elles qui jouent un rôle dans l'intersubjectivité et la dynamique communautaire.
19. Cette distinction renvoie à la distinction entre personne et personnage [DOU 03].

| | Page de profil | Journal | Autres journaux | Identité personnelle | Identifiants | Identité diégétique | Autre | Déclaratif | Calculé par le système |
|---|---|---|---|---|---|---|---|---|---|
| Identifiant | X | | | | X | | X | | |
| numéro de compte | X | | | | X | | | | X |
| Pseudonyme | X | X | (X) | | X | | X | | |
| Icône | X | X | | | X | | X | | |
| Titre du journal | X | X | | | X | | X | | |
| Type de compte | X | | | X | x | | X | | |
| Date de création | X | | | X | x | | | | X |
| Age, sexe, ville | X | | | X | | | X | | |
| Site personnel en lien | X | X | | | | X | X | | |
| Contacts de messagerie | X | | | | | X | X | | |
| Biographie | X | | | x | X | | X | | |
| Liste indexée de centres d'intérêt | X | | | X | X | | X | | |
| Date de la dernière mise à jour | X | | | X | x | | | | X |
| Nombre d'entrées | X | | | X | X | | | | X |
| Liste d'amis | X | | | x | X | | X | | |
| Amis en commun | X | | | | X | | | | X |
| Communautés | X | | | | X | | X | | |
| Nombre de commentaires reçus et envoyés | X | | | X | X | | | | X |
| Mise en page | | X | | X | X | | X | | |
| Page de publication personnelle | | X | | | X | | X | | |
| Calendriers | | X | | X | X | | | | X |
| Posts datés | | X | (X) | X | X | | X | | |
| Datation des posts | | X | (X) | X | (X) | | | | X |
| Page des favoris syndiqués | | X | | | X | | | | X |
| Commentaires envoyés | | X | X | X | X | | X | | |

**Figure 1** *Classification des éléments de la représentation de soi*[20] *dans Livejournal.*

---

20. La page de profil contient les identifiants (pseudo, numéro de compte), des informations sur l'utilisateur comme auteur du blog (icone, nom de l'auteur, nom du journal, le type de compte, la date de création,), des informations sur la personne (âge, sexe, ville, site personnel, numéros de messageries instantanées), des informations faisant le lien entre l'un et l'autre (biographie, liste indexée de centres d'intérêt en saisie libre), des informations sur l'activité de La difficulté survenant lors de la tentative de classification des signes du point de vue des identités est leur intrication. Univers diégétique et monde réel sont étroitement liés par le type de compte, la date de création du journal, les mises à jour ou encore, dans une moindre mesure, les centres d'intérêt. Il est difficile d'apprécier la segmentation de la représentation, néanmoins elle révèle quels sont les signes qui, proches du réel, peuvent donner la mesure de l'implication personnelle de l'auteur et de son immersion personnelle en le *soi* de la représentation.

### *2.2 Intimité et porosité de la représentation*

Le niveau de confidentialité des posts publiés est paramétrable. « Publics », les posts sont visibles aux membres et non membres, « semi-privés » par les favoris uniquement, et lorsqu'ils sont « privés », seul l'auteur, administrateur du blog, peut les consulter. Ainsi, un journal publié intégralement en mode semi-privé ou privé est inaccessible à l'utilisateur lambda sans qu'il puisse toutefois être certain que le diariste soit inactif[21]. Cette ambiguïté dessine une forme d'intimité de la représentation. En réservant des données personnelles à un nombre limité d'internautes, l'usager sculpte les contours de la structure sociale de son identité diégétique.

Jusqu'à présent, nous avons développé une conception de la représentation de soi rassemblant les éléments qui manifestent une interaction de l'utilisateur avec l'environnement personnel et intersubjectif. Nous avons vu qu'un certain nombre d'informations disponibles étaient produites automatiquement par le système et qu'elles manifestaient indirectement l'identité de l'utilisateur, tant personnelle que diégétique. Restent encore les informations envoyées par les autres utilisateurs, à savoir les commentaires qui font partie de la représentation du favori dans le blog duquel ils sont syndiqués. L'utilisateur peut choisir d'en recevoir du tout venant, de ses amis seulement ou refuser tout commentaire, gérant ainsi finement la porosité de sa représentation.

---

mise à jour (date de la dernière mise à jour, nombre d'entrées) et l'activité communautaire (liste des amis, amis en commun, communautés, le nombre de commentaires reçus et envoyés).

Le journal se constitue de plusieurs onglets : page de publication personnelle, la zone de syndication avec les favoris, calendriers, page rassemblant des passages mémorables et un lien vers le site personnel. La page de publication contient l'icône de l'utilisateur, son pseudo, le titre du journal, les posts datés, les commentaires. Il est mis en page, à la différence de la page de profil qui n'est pas personnalisable.

21. Ce peut être même une stratégie d'incitation à devenir favori, comme peut le sous-entendre cette diariste qui précise sur sa page de profil que la majorité des posts ne sont ouverts qu'aux amis : « P.S. La majeure partie des entrées sont en mode 'friends only' » Cf. http://www.livejournal.com/userinfo.bml?user=comtesse_noire

*2.3 Contrôler l'œil du système*

Lors de son inscription, le diariste renseigne des champs dont le contenu, partiellement affiché dans l'espace public de la représentation, est ipso facto déclaratif. Les deux dernières colonnes de la figure 1 révèlent qu'en outre, près d'un tiers des éléments disponibles sont calculés par le système. Sur les 7 éléments calculés, 3 sont relatifs à l'implication en termes de fréquence (nombre d'entrées, de commentaires, calendriers) et de proximité temporelle au moment de la consultation (date de la dernière mise à jour), 2 sont relatifs à l'implication communautaire en terme d'expression des affinités (les blogs favoris sont automatiquement syndiqués dans la zone dédiée) et de réciprocité (noms des favoris ayant également placé l'utilisateur en favori) et enfin, la datation même des posts, qui manifeste le moment où l'utilisateur, comme personne, est connecté, n'est pas modifiable ni suppressible. Certains de ces signes ne peuvent être modifiés (numéro de compte, date de création), tandis que la plupart peuvent être contrôlés indirectement par l'utilisateur. Mettre à jour fréquemment son blog, ajouter les blogs lus à la liste des favoris, poster des commentaires sont les moyens d'influer sur ces signes, faussement involontaires, de la fréquence, de la proximité temporelle au moment de la consultation et de l'implication communautaire. Du point de vue ergonomique, le calcul et l'affichage des ces informations par le système sont des indicateurs commodes qui renseignent rapidement le lecteur sur l'actualité de sa lecture et l'aura relationnelle de son auteur. De ce désir de lire des pages les plus fraîchement écrites naît la nécessité pour soi-même de répondre à ses propres attentes.

## 3. La représentation de soi dans le regard de l'alter ego

*3.1 Portrait de l'auteur-lecteur*

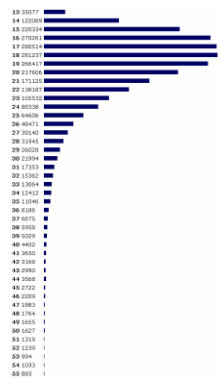

**Figure 2:** *Répartition des diaristes de Livejournal par leur âge*[22].

Contrairement aux communautés de rencontre ou de notation, les blogs ont cette particularité de drainer un large public. Comme le montre la figure 2, les plus jeunes utilisateurs ont 13 ans, les plus âgés 55. Le coeur de cible se situe entre 16 et 19 ans.

Des entretiens exploratoires[23], il ressort qu'au début de leur pratique, même s'ils ne cherchaient initialement qu'à expérimenter les outils d'expression, les utilisateurs ont découvert les fonctionnalités communautaires et s'en sont emparés, « se prenant au jeu », pour accroître leur audience et entrer dans le jeu communautaire. L'intérêt pour la participation communautaire a été cité comme une motivation survenue après manipulation, même si par ailleurs les témoins avaient une pratique communautaire élevée dans d'autres services.

La découverte du blog d'une connaissance ou d'une célébrité a été citée comme facteur de l'adoption d'un service particulier : de la lecture à l'inscription il semble qu'il n'y ait qu'un pas. Avoir un site personnel, donner un cadre à son écriture sont deux motifs cités parmi les motivations initiales. L'usage du blog est perçu comme relevant d'une simple curiosité pour certains, et de la recherche d'une discipline d'écriture pour d'autres. Entretenir son lectorat nécessite une fréquence quotidienne de publication ; c'est la raison pour laquelle nombre d'usagers ne s'y risquent pas. Comme le résume cet utilisateur (homme, 29 ans, pratiquant la messagerie instantanée au moins une heure par jour et lecteur quotidien de blogs BD), « pour écrire quotidiennement, il faut fournir du contenu, avoir quelque chose à dire. Je n'ai pas le temps et je n'ai pas le contenu ».

La figure 3 montre que le pourcentage de blogs actualisés est proportionnellement faible. Deux tiers des blogs ont été mis à jour au moins une fois : le tiers restant montre que les tentatives velléitaires sont nombreuses. Un quart l'ont été pour la dernière fois dans le mois. Or, comme nous l'avons vu, la régularité de la publication n'est pas seulement une question d'envie, elle est nécessaire pour exister au sein de l'univers diégétique. Ne pas publier est aussi priver ses lecteurs d'un espace de co-construction du sens [DES 03]. Espace d'expression personnelle, il est aussi un espace d'expression communautaire.

---

22. Source : ces statistiques ont été relevées sur le site de *Livejournal* le 23 mars 2005. Cf. : http://www.livejournal.com/stats.bml

23. Ces entretiens semi-ouverts s'inscrivent dans l'exploration plus large des usages des outils de communication à distance. Pour l'année 2004, ils ont concerné une vingtaine d'utilisateurs français homme et femme entre 25 et 40 ans. Dans cet échantillon non représentatif, les utilisateurs des blogs sont minoritaires, mais il est intéressant également du point de vue de l'étude des représentations, d'écouter les raisons pour lesquelles les usagers n'utilisent pas de blog.

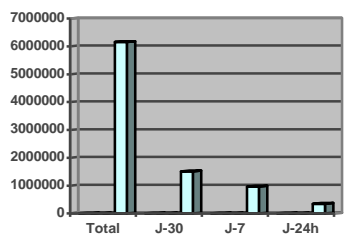

**Figure 3** *Nombre de mises à jour à un mois, une semaine et un jour*[24].

### *3.2 Lire et être lu : les clés d'un succès partagé*

Les diaristes peuvent faire connaissance en consultant les favoris de leurs favoris et en postant des commentaires. Tout utilisateur placé en favori en est informé par la liste correspondante du panneau de profil qui figure également dans une rubrique du back-office. Mettre quelqu'un en favori est une forme de reconnaissance qui manifeste que le blog en question est apprécié et lu ; cette marque de reconnaissance peut impliquer une attention en retour. Poster des commentaires est une autre façon de diffuser sa représentation personnelle et de l'étayer dans un contexte de dialogue et d'échange. Les moteurs et techniques de recherche sont une autre façon de découvrir des membres. Les expérimenter n'est pas seulement apprendre à saisir les bons critères pour trouver un interlocuteur mais aussi savoir comment on est trouvé : ce double aspect nécessite un double examen[25].

Les recherches disponibles avec un compte gratuit sont la recherche aléatoire et par centre d'intérêt. Les fonctionnalités avancées donnent accès à un moteur de recherche multicritères.

- La recherche aléatoire permet de découvrir un nouveau blog d'un simple clic, mais son principal inconvénient est que la pertinence de la mise en relation est indéterminée.

- La recherche par centre d'intérêt est accessible soit en cliquant sur l'onglet de recherche correspondant dans le menu, soit par le profil utilisateur (en cliquant sur un de ses centres d'intérêt). Les résultats se présentent sous forme d'une liste d'utilisateurs possédant un centre d'intérêt commun. Pour être mieux référencés, les membres rentrent plus d'une cinquantaine de mots-clé. Cette fonctionnalité a pour principal inconvénient d'être l'objet de détournements qui rendent sa pertinence superficielle, comme le copier coller des listes d'autres utilisateurs. Un autre inconvénient est qu'elle ne prend en compte qu'un seul centre d'intérêt à la fois, et qu'il n'est pas possible de classer les résultats par ordre de pertinence, en mettant en valeur les utilisateurs qui ont le plus d'affinités.

---

24. Cf. note 22.

25. Cette quête des règles sous-jacentes à la conception des services n'est pas spécifique au blog. Une utilisatrice de Meetic (F, 38 ans) explique qu'elle pratiquera ce service jusqu'à ce qu'elle ait compris son fonctionnement, la « leçon » qui le sous-tend.

- Le moteur de recherche multicritère croise localisation géographique (pays, état, ville[26]), date ou période de la mise à jour, âge, centres d'intérêt, amis communs. Ces informations, qui relèvent de l'identité *personnelle* (Cf. figure 1), sont proches de celle qui sont délivrées dans les sites de rencontre ou les outils communautaires comme ICQ[27], à cette différence près qu'elles ne font pas intervenir le critère physique[28]. Malgré cela, les fonctionnalités de recherche avancées semblent valoriser l'identité *personnelle* au détriment de l'identité *diégétique*.

L'insuffisance des recherches disponibles parmi les fonctionnalités basiques de Livejournal valorise l'offre avancée[29], qui se révèle finalement décevante pour l'utilisateur qui chercherait à découvrir des diaristes avec lesquels il partage des affinités relevant de la diégèse.

### 3.3 Affinités et détournements

A côté des moteurs de recherche peu appropriés, les fonctionnalités communautaires restent encore les plus pertinentes. Cette soumission de la visibilité à l'inscription de soi comme favori dans les listes des autres diaristes pose un dilemme entre le désir d'intimité et le désir d'être lu. Cette appréhension du dévoilement de soi peut être franchie par l'adoption de la stratégie d'automédiation comme un jeu (H, 37 ans). Cette stratégie consisterait à poster des commentaires pour séduire et être visible, attirer le lecteur sur sa page et peut-être bénéficier de sa sympathie et de sa reconnaissance par l'inscription. Comme l'exprime ce diariste (H, 37 ans), usager assidu de tout type d'outil de communication par ordinateur depuis

---

26. La localisation géographique est fixe là où un moblog pourrait proposer une géolocalisation. La recherche par région permet d'accéder aux blogs via une carte des Etats Unis (par état) et une liste de pays. La moitié des utilisateurs sont en effet américains. Les autres pays les plus représentés sont le Canada (200 000 comptes), l'Angleterre (163 000) et l'URSS (127 000). La France rassemble 2500 utilisateurs. Même s'il existe une véritable communauté francophone liée par les favoris, la plupart des journaux sont anglophones. De nombreux francophones rédigent leur journal en anglais ou dans les deux langues.
27. ICQ est une messagerie instantanée avec des fonctionnalités (moteurs de recherche) et des services communautaires (jeux). http://www.icq.com
28. Ceci dit, beaucoup d'utilisateurs utilisent une photo d'identité comme icone personnel.
29. 98,4 % des utilisateurs ont un compte gratuit et ne bénéficient donc pas des fonctionnalités avancées. Cependant, il est intéressant d'observer comment le logiciel construit l'identité numérique de ses membres et dispense les avantages dans les comptes payants.

15 ans et pratiquant le blog français de 20six depuis 6 mois, cette stratégie de l'intersubjectivité n'est pas sentie comme étant utilitariste. Il lit avec plaisir ses blogs favoris et les marques de reconnaissance qu'il dispense sous forme de bonbons[30] sont sincères[31].

Livejournal ne propose ni de statistiques en page d'accueil incitant à obtenir le plus de récompenses (20six), ni de système de navigation efficace qui permette aux membres de se connaître autrement qu'en explorant les réseaux de favoris et en jouant le jeu de la syndication et de la publication de commentaires. L'examen du logiciel révèle que les détournements sont accessibles, aisés et ludiques. Au delà du développement d'une stratégie d'automédiation reposant sur une maîtrise active des fonctionnalités, la pratique disciplinée et prolongée du blog conduit à se poser la question des motivations du lecteur. Parmi les solutions trouvées par les personnes interrogées, le motif de la sincérité est récurent [BOY 05]. Sincérité avec soi-même, sincérité avec l'autre font des blogs des espaces intimes où des représentations se mêlent dans la confidence, où la réciprocité des liens traduit une bienveillance partagée.

**4. Visualisation et construction**

Le module Touchgraph (figure 4) présente un champ de saisie textuel. Seuls les pseudos d'utilisateurs peuvent y être rentrés, non les centres d'intérêt. Des relations *un-à-plusieurs* et plusieurs-à-un peuvent être représentées simultanément. Un nombre paramétrable de liens peut entrer ou sortir de chaque nœud. En cliquant sur l'icone d'un utilisateur, une fenêtre de navigateur s'ouvre et affiche sa feuille de profil ; en cliquant sur un centre d'intérêt, le navigateur affiche tous les utilisateurs qui le partagent parmi ceux dont la représentation a été préalablement déployée.

---

30. 20six intègre à ses fonctionnalités communautaires les bonbons. Chaque utilisateur reçoit chaque jour du système 5 bonbons qu'il peut dispenser à qui bon lui semble, même à lui-même. Sur la page d'accueil du service correspond un classement des utilisateurs qui en ont reçu le plus.

31. Cet utilisateur avoue qu'il a trouvé un autre moyen de tricher, en se donnant à lui-même tous ses bonbons dès la mise à zéro des statistiques de la page d'accueil. Ainsi promu pendant quelques minutes au premier rang des utilisateurs ayant le plus de bonbons, il bénéficie d'une forte visibilité et agrandit très nettement son lectorat. Cependant, la procédure révélant un machiavélisme bien peu compatible avec l'esprit officiel du service, il se garde de l'appliquer trop souvent par peur d'être repéré par l'un de ses nombreux lecteurs.

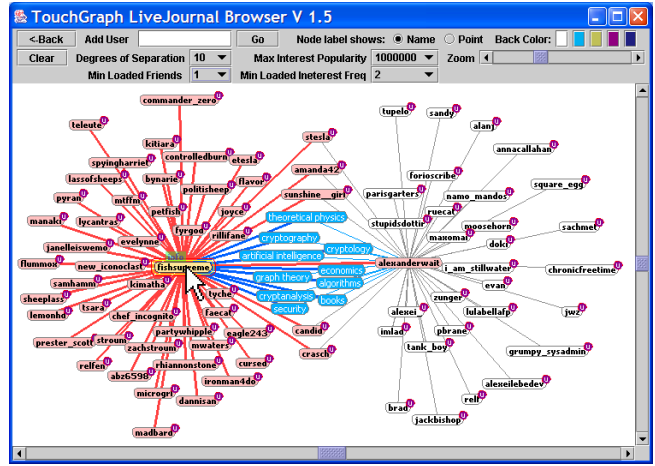

**Figure 4** *Visualisation de deux représentations de diaristes dans Touchgraph. Les items rouges représentent des utilisateurs, les items bleus les centres d'intérêt.*

Touchgraph semble *a priori* d'un intérêt limité dans la dynamique communautaire car il passe sous silence la plupart des informations ayant pour fonction la mise en relation au bénéfice de considérations statistiques comme le filtrage des centres d'intérêts les plus fréquents. Son point fort réside en ce qu'il est une conversion graphique et manipulable de listes que la surabondance rend difficilement consultables dans les pages de profil textuelles. Il ne parvient cependant pas à relever le pari de la simplification. Les items apparaissent initialement superposés. La signification d'un tel graphe dépend d'un arrangement spatial pertinent, or le système de coordonnées planaires n'a pas de signification. L'utilisateur en dénoue les nœuds et les organise. Un clic sur un centre d'intérêt le lie à ses autres occurrences[32].

Là où le logiciel est défaillant se libèrent l'initiative et la créativité. Comme il le ferait dans un logiciel de mindmapping, l'usager organise les centres d'intérêt et les liens sociaux, donnant lui-même une signification à cette spatialisation. Il prend ainsi conscience des relations entre des éléments auparavant dissociés dans la linéarité sans motif des longues listes de Liveblog. Cette mise en forme manuelle peut avoir une réelle plus-value pour le développement de la conscience de soi et de l'autre.

---

32. Par contre, si un centre d'intérêt est commun à un utilisateur et à des utilisateurs représentés en lien, il n'est pas mis en commun.

## 4. Conclusion

L'analyse sémiopragmatique de la représentation de l'utilisateur dans Livejournal a permis d'identifier les signes de l'interface qui participent à l'identité, l'identification et à la mise en relation des utilisateurs. L'approche de la représentation comme système permet d'apprécier les écarts et les constantes d'appropriation dans la construction une représentation personnelle ou communautaire. A rebours de sa parenté avec les logiciels de networking social, Touchgraph, utilisant les informations personnelles trop superficiellement, n'apporte à cet égard pas de réelle plus-value communautaire mais peut fournir un outil intéressant de réflexion sur les réseaux sociaux, nourrissant un intérêt manifeste des utilisateurs pour l'observation sociologique[33], son usage se rapprochant par là-même davantage des logiciels de *mindmapping* que des logiciels de *networking social*. Si les outils sont mal adaptés à la situation de communication propre au cyberjournal, cette combinaison d'un blog à un logiciel de visualisation manifeste une tendance à donner une forme visuelle simple à ce qui est profondément complexe, l'expression de soi.

La représentation de l'usager est un cadre auquel s'arrime la perception et se façonne la subjectivité [WEIS 99]. En deçà de l'euphorie de la reconnaissance mutuelle [DES 03], de l'intimité partagée, et en deçà d'un narcissisme avide de reconnaissance [ALL 03], le logiciel trace ses règles et cadres d'usage. Par la création de pages personnelles et de blogs, l'individu serait amené à promouvoir une *authenticité réflexive*[34] dans une prétention *quasi-artistique* [ALL 03] soumise à un jugement réfléchissant. Les fonctionnalités communautaires de Livejournal donnent corps à ce jugement, soumettant la visibilité de soi à son évaluation par les favoris. La crainte d'être lu fait place au désir de s'exposer tout en restant dans une ombre rassurante [CAU 03]. La stratégie d'automédiation, déterminée par le logiciel, consisterait alors en une séduction de l'alter ego pour obtenir sa modeste approbation. Or la majorité des diaristes n'ont pas conscience de raconter une histoire et considèrent leur blog comme l'expression même de leur personnalité. La recherche de l'approbation pourrait paraître risquée.

La majeure partie des diaristes interrogés créent des blogs, y écrivent régulièrement et en changent. Ce changement s'accompagne parfois d'une mutation de l'identité visuelle et diégétique, comme si patiemment, à tâtons, ils tentaient de se détacher du déterminisme identitaire de la permanence pour recommencer à se déployer légèrement plus loin, se rapprochant peu à peu d'une vérité de soi-même encore fragile que seuls ces petits mondes pourraient accueillir.

---

33. La plupart des personnes interrogées (cf. n.23) estiment que les utilisateurs sont un échantillon représentatif de la société.
34. Expression de Ferrara A., Reflexive Authenticity. Rethinking the Project of Modernity, Londres, Toutledge, 1990. Cité par [ALL 03] p. 18, n. 37.

## 5. Bibliographie